\documentclass[12pt]{article}

\usepackage{amssymb}
\usepackage{latexsym}

\usepackage{epsfig}
\usepackage{graphicx}

\usepackage{amsmath,amsthm,amsfonts,amssymb}

\newcommand{\be}{\begin{equation}}
\newcommand{\ee}{\end{equation}}
\newcommand{\bea}{\begin{eqnarray}}
\newcommand{\nn}{\nonumber}
\newcommand{\eea}{\end{eqnarray}}

\begin{document}
%%%%%%%%%%%%%%%%%%%%%%%%%%%%%%%%%%%%%%%%%%%%%%
%\begin{figure}[h]
%\centering
%\includegraphics[scale=0.5]{a:figure1p.ps}
%\end{figure}
%%%%%%%%%%%%%%%%%%%%%%%%%%%%%%%%%%%%%%%%%%%%%%

\begin{titlepage}
\begin{flushright}

NTUA--103--2004

gr-qc/0403090\\

\end{flushright}

\begin{centering}
\vspace{.41in}
{\large {\bf Chaotic Inflation on the Brane with Induced Gravity}}\\

\vspace{.4in}

 {\bf E.~Papantonopoulos}$^{a}$ and {\bf V.~Zamarias} $^{b}$ \\
\vspace{.2in}

 National Technical University of Athens, Physics
Department, Zografou Campus, GR 157 80, Athens, Greece. \\

\end{centering}
\vspace{1.0in}

\begin{abstract}

We study the slow-roll inflationary dynamics in a self-gravitating
induced gravity braneworld model with bulk cosmological constant.
For $E\gg M_{5}^{3}/M^{2}_{4}$ we find important corrections to
the four-dimensional Friedmann equation which bring the standard
chaotic inflationary scenario in closer agreement with recent
observations. For $M^{3}_{5}/M^{2}_{Pl}\ll E \ll
M^{3}_{5}/M^{2}_{4}$ we find five-dimensional corrections to the
Friedmann equation, which give the known Randall-Sundrum results
of the inflationary parameters.

\end{abstract}

\vspace{1.0in}
\begin{flushleft}

$^{a}$ e-mail address:lpapa@central.ntua.gr \\$ ^{b}$ e-mail
address:zamarias@central.ntua.gr

\end{flushleft}

\end{titlepage}

\section{Introduction}

The recent observation data from the Wilkinson Microwave
Anisotropy Probe (WMAP) \cite{wmap}, show strong support for the
standard inflationary predictions of a flat Universe with
adiabatic density perturbations in agreement with the simplest
class of inflationary models \cite{inflation}. In particular,
these observations had significantly narrowed the parameter space
of slow-roll inflationary models. We are entering an area where
the physics in the early universe can be probed by upcoming
high-precision observational data.

In light of these developments, it is important to understand
further the inflationary scenario from the theoretical point of
view and also from a more phenomenological approach. A new idea
that was put forward is, that our universe lies in a
three-dimensional brane within a higher-dimensional bulk spacetime
and this idea may have important consequences to our early time
Universe cosmology. The most common scenario, is the
Randall-Sundrum model of a single brane in an AdS bulk
\cite{randall}. In this model the Friedmann equation is modified
in early times while standard cosmology is recovered at late times
\cite{binetruy,csaki}. The slow-roll inflation was applied to this
model \cite{maartens}, and it was found that the extra terms in
the Friedmann equation act as friction terms which damp the
rolling of the scalar field, allowing steeper potentials. However,
the recent observational data from CMB anisotropies, seems to
exclude steep potentials and strongly constrain monomial
potentials in the Randall-Sundrum scenario \cite{shinji}.

We can also study inflation in other braneworld cosmological
models. These models are mainly generalizations of the
Randall-Sundrum model. The most interesting models are those which
are based on the generalization of the Randall-Sundrum
gravitational action. The gravitational action can be generalized
in two ways. The first is to add a four-dimensional scalar
curvature term in the brane action. This induced gravity
correction arises because the localized matter fields on the
brane, which couple to bulk gravitons, can generate via quantum
loops a localized four-dimensional world-volume kinetic term for
gravitons~\cite{hc,dvali1}. The second is a Gauss-Bonnet
correction to the five-dimensional action. This gives the most
general action with second-order field equations in five
dimensions~\cite{lovelock}.

The induced gravity model, with no brane tension (i.e., the brane
is not self-gravitating) and no bulk cosmological constant, leads
to very different behaviour than the Randall-Sundrum model. In the
Randall-Sundrum case, gravity becomes five-dimensional at high
energies, so that general relativity is modified in the early
universe. By contrast, in the induced gravity case, corrections to
general relativity become significant at low energies/ late times,
and the early universe evolution agrees with standard general
relativity~\cite{dvali1,indgrav,deffayet}. Astrophysical
implications of induced gravity have also been
considered~\cite{astro}.

The Gauss-Bonnet model is like the Randall-Sundrum model in the
sense that modifications to general relativity arise in the early
universe at high energies. The graviton zero mode is also
localized at low energies, as in the Randall-Sundrum
case~\cite{gbloc}. The brane cosmology of the Gauss-Bonnet theory
has been investigated in
Refs.~\cite{dl,germani,charmousis,gbmore}. In \cite{lidnun} the
slow-roll inflation was applied to the Randall-Sundrum model with
a Gauss-Bonnet correction term and it was found that if the
inflation is driven by an exponential inflaton field, the
Gauss-Bonnet term allows for a spectral index of the scalar
perturbation spectrum to take values closer to the recent
observational data.

Recently the cosmology of the Randall-Sundrum braneworld was
studied included both curvature correction terms: a
four-dimensional scalar curvature from induced gravity on the
brane, and a five-dimensional Gauss-Bonnet curvature term
\cite{papa}. The combined effect of these curvature corrections to
the action removes the infinite-density big bang singularity,
although the curvature can still diverge for some parameter
values. A radiation brane undergoes accelerated expansion near the
minimal scale factor, for a range of parameters. At late times,
conventional cosmology is recovered.

In the induced gravity model when the brane tension and bulk
cosmological constant are included, the late-time modifications
persist, with less fine-tuning needed, but further modifications
are introduced ~\cite{hc,ktt,sahni,burin,mmt}. In this work we
show that the inclusion of the brane tension and bulk cosmological
constant, have also important consequences in the early time
evolution. For $E\gg M_{5}^{3}/M^{2}_{4}$ we find significant
corrections to the standard four-dimensional Friedmann equation.
For $M^{3}_{5}/M^{2}_{Pl}\ll E \ll M^{3}_{5}/M^{2}_{4}$ we find
Randall-Sundrum type five-dimensional corrections to the Friedmann
equation for an AdS or Minkowski bulk as expected. Applying the
slow-roll inflationary formalism in the case of a quadratic
potential, we find that the correction terms to the standard
Friedmann equation, bring the standard chaotic inflationary
scenario closer to the observations. We also find that because
gravity becomes stronger in the high energy regime, chaotic
inflation can occur for values of the inflaton field below the
Planck scale.

The paper is organized as follows. In Sec. II, we analyse the
cosmological evolution of a self-gravitating braneworld model with
induced gravity where a cosmological constant is also included,
and we find all possible corrections, in the allowed parametric
space, to the Friedmann equation. In Sec. III, we apply the
slow-roll inflationary formalism with a quadratic potential to the
resulted modified Friedmann equations. In Sec. IV, we discuss
chaotic inflationary models allowed for various choices of
parameters, and Sec. V gives our conclusions.

\section{Cosmological evolution of a brane-universe with induced gravity}

The effective Einstein equations on the brane, when a
four-dimensional scalar curvature term is included in the bulk
action, are \cite{mmt,kofinas} \bea
\Big(1+\frac{\lambda}{6}r_{c}\kappa_{5}^2\Big)G^{\mu}_{\nu}&=&\frac{\lambda}{6}
 \kappa_{5}^4 T^{\mu}_{\nu}-r_{c}\kappa_{5}^2 \mathcal{K}^{\mu}_{
 \nu\rho\sigma}(T_{\alpha \beta})G^{\rho \sigma}
-\frac{1}{2}\Big(\Lambda_{5}+\frac{\lambda^{2}}{6}\kappa_{5}^{4}\Big)
\delta^{\mu}_{\nu} \\ \nn &+& \kappa^{4}_{5} \pi^{(T)\mu}_{\nu}+
r_{c}^2 \pi^{(G)\mu}_{\nu}-\mathcal{E}^{\mu}_{\nu}\,,\label{torii}
\eea where \bea \mathcal{K}^{\mu}_{\nu \rho \sigma}&=& \frac{1}{4}\Big{(}\delta^{\mu}_{\nu} T_{\rho \sigma}
-\delta^{\mu}_{\rho} T_{\nu \sigma}-\delta_{\nu \sigma} T^{\mu}_{\rho}\Big{)}\\
&+&\frac{1}{12}\Big{(} T^{\mu}_{\nu}\delta_{\rho \sigma}+\Big{(}
T^{\lambda}_{\lambda} \Big{)}\Big{(}\delta^{\mu}_{\rho}\delta_{\nu
\sigma }-\delta^{\mu}_{\nu}\delta_{\rho \sigma } \Big{)}
\Big{)},\label{kapa}\\
\pi^{(T)\mu}_{\nu}&=&-\frac{1}{4}T^{\mu}_{\lambda}T^{\lambda}_{\nu}+\frac{1}{12}T^{\lambda}_{\lambda}T^{\mu}_{\nu}+
\frac{1}{8}T^{\lambda}_{\rho}T^{\rho}_{\lambda}\delta^{\mu}_{\nu}-\frac{1}{24}(T^{\lambda}_{\lambda})^{2}
\delta^{\mu}_{\nu}, \label{pit} \\
\pi^{(G)\mu}_{\nu}&=&-\frac{1}{4}G^{\mu}_{\lambda}G^{\lambda}_{\nu}+\frac{1}{12}G^{\lambda}_{\lambda}G^{\mu}_{\nu}+
\frac{1}{8}G^{\lambda}_{\rho}G^{\rho}_{\lambda}\delta^{\mu}_{\nu}-\frac{1}{24}(G^{\lambda}_{\lambda})^{2}
\delta^{\mu}_{\nu}. \label{pig} \eea The induced-gravity crossover
length scale $r_{c}$ is
 \bea
r_{c}=\frac{\kappa_5^2}{\kappa_4^2}=\frac{M_4^2}{M_5^3}\,,
 \label{distancescale}
 \eea
the fundamental ($M_5$) and the four-dimensional ($M_4$) masses
are given by
 \bea
\kappa_{5}^{2}=8\pi G_{5}=M_{5}^{-3}\,,~~ \kappa_{4}^{2}=8\pi
G_{4}=M_{4}^{-2}\,. \label{planck}
 \eea

 In a cosmological setup in which the projected metric on the
 brane is a spatially flat Friedmann-Robertson-Walker model, with
 scale factor $a(t)$, the Friedmann equation on the brane is
 \cite{indgrav,deffayet}
\bea H^{2}&=&\frac{\kappa_{4}^{2}}{3}(\rho + \lambda)
+\frac{{2}}{r_{c}^2}-\frac{k}{a ^{2}}\nn\\&+&~{} \epsilon\,\, \!
\frac{1}{\sqrt{3}\,r_{c}}\left[{ 4\kappa_{4}^{2}(\rho+\lambda) -
2\Lambda_{5}+{12\over r_{c}^2}- \frac{12\mathcal{C}}{a
^{4}}}\right]^{1/2}\!,  \label{igr}
 \eea
where $\mathcal{C}$ is an integration constant arising from the
Weyl tensor $\mathcal{E}^{\mu}_{\nu}$ and $\epsilon= \pm 1$.
 The second Friedmann equation
can also be calculated  \bea
\dot{H}&=&-\frac{\kappa_{4}^{2}}{2}\rho(1+w)+\frac{k}{a ^{2}}\nn
\\&-&\epsilon\,\, \!
\frac{1}{\sqrt{3}\,r_{c}}\frac{3\kappa_{4}^{2}\rho(1+w)-\frac{12\mathcal{C}}{a
^{4}}}{\left[{ 4\kappa_{4}^{2}(\rho+\lambda) -
2\Lambda_{5}+{12\over r_{c}^2}-\frac{12\mathcal{C}}{a ^{4}}
}\right]^{1/2}\!}\,\,, \label{secondfried} \eea where $w$ is given
by the equation of state $p=w\rho$.

 Equation
(\ref{igr}) describes the cosmological evolution of the
brane-universe with induced gravity. The fundamental parameters
appearing in (\ref{igr}) and (\ref{secondfried}) are: two energy
scales, i.e. the fundamental Planck mass $M_{5}$, and the induced
gravity crossover energy scale $r^{-1}_{c}$, and two vacuum
energies, i.e. the bulk cosmological constant $\Lambda_{5}$ and
the brane tension $\lambda$. In this work we are mainly interested
on the inflationary dynamics driven by a scalar field $\phi$ with
a self-interaction potential $V(\phi)$. Therefore, we will put the
effective cosmological constant equal to zero. The effective
cosmological constant can be calculated by taking the
$a\rightarrow\infty$ limit of (\ref{igr}) \be \Lambda_{\rm
eff}=\kappa_{4}^{2}\lambda+\frac{6}{r^{2}_{c}}+\epsilon
\frac{\sqrt{6}}{r^{2}_{c}}\sqrt{\left(2\kappa_{4}^{2}\lambda-
\Lambda_{5} \right)r^{2}_{c}+6}\,\,,\label{lambdaIG} \ee and
putting it to zero we get,
 \be
\Lambda_{5}=-\frac{1}{6}r^{2}_{c}\kappa^{4}_{4}\lambda^{2}.
\label{finetune} \ee For $\epsilon=-1$ the brane tension can be
assume to be positive, while for $\epsilon=+1$ the brane tension
is negative. We will discuss the implications of a negative brane
tension in the following.
 In the case of a Minkowski bulk
($\Lambda_{5}=0$) we cannot put the $\Lambda_{eff}=0$, because the
induced gravity decouples, and we find the Randall-Sundrum model.
However, setting only $\Lambda_{5}=0$, the brane tension can be
assumed to be positive for both signs of $\epsilon$.

For $\Lambda_{eff}=0$ using (\ref{finetune})  and setting $k$ and
$\mathcal{C}$ equal to zero, equation (\ref{igr}) becomes \be
H^{2}=\frac{\kappa^{2}_{4}\rho}{3} \Big{\{}
1+\frac{6}{\kappa^{2}_{4}
r^{2}_{c}\rho}+\frac{\lambda}{\rho}+\epsilon\,
\frac{2\sqrt{3}}{\kappa_{4} r_{c}\rho^{1/2}}\Big{(}
1+\frac{3}{\kappa^{2}_{4} r^{2}_{c}\rho}+\frac{\lambda}{\rho} +
\frac{\lambda}{\rho} \frac{\kappa^{2}_{4} r^{2}_{c}\lambda}{12}
\Big{)}^{1/2}\Big{\}}. \label{approxfried} \ee To analyse the
above Friedmann equation we define the dimensionless parameter \be
\mu=\frac{\kappa^{2}_{4}r^{2}_{c}\lambda}{6}, \label{mu} \ee and we distinguish the following cases:\\
\\
1)\,\,\underline{$\mu \gg 1$}.\, In the high energy limit
$\rho\gg\lambda$, we have $\kappa_{4}^{2}r_{c}^{2}\rho\gg 1$ and
the Friedmann equation (\ref{approxfried}) is approximated by \bea
\kappa^{2}_{4}r^{2}_{c}\lambda \Big{(}\frac{\lambda}{\rho}
\Big{)}\ll1, &H^{2}&=\frac{\kappa^{2}_{4}\rho}{3} \Big{(}
1+\epsilon \frac{2\sqrt{3}}{\kappa_{4} r_{c}\rho^{1/2}}\Big{)},\label{approxm1} \\
\kappa^{2}_{4}r^{2}_{c}\lambda \Big{(}\frac{\lambda}{\rho}
\Big{)}\sim 1,  &H^{2}&=\frac{\kappa^{2}_{4}\rho}{3} \Big{\{}
1+\epsilon  \frac{2\sqrt{3}}{\kappa_{4} r_{c}\rho^{1/2}}\Big{(} 1+
\frac{\kappa^{2}_{4}
r^{2}_{c}\lambda^{2}}{12\rho} \Big{)}^{\frac{1}{2}}\Big{\}} \label{approxm2}\\
\kappa^{2}_{4}r^{2}_{c}\lambda \Big{(}\frac{\lambda}{\rho}
\Big{)}\gg1,\epsilon=+1, &H^{2}&=\frac{\kappa^{2}_{4}\rho}{3}
\Big{(} 1+2 \frac{\lambda}{\rho}\Big{)},\label{casee+}\\
\epsilon=-1, &H^{2}&=\frac{\kappa^{2}_{4}\rho}{3} \Big{(}
1-\frac{6}{\kappa^{2}_{4} r^{2}_{c}\lambda}\Big{)}\label{casee-}.
\eea In the low energy limit $\rho\ll\lambda$, (\ref{approxfried})
is approximated by \bea \epsilon=+1,\,\,
\kappa_{4}^{2}r_{c}^{2}\rho\gg 1,\,\,
H^{2}&=&\frac{\kappa^{2}_{4}\rho}{3} \Big{(} 1+2
\frac{\lambda}{\rho}\Big{)},\\
\kappa_{4}^{2}r_{c}^{2}\rho\ll 1,\,\,
H^{2}&=&2\frac{\kappa^{2}_{4}\lambda}{3} \Big{(} 1+
\frac{6}{\kappa^{2}_{4} r^{2}_{c}\lambda}\Big{)},\\
\epsilon=-1,\quad \quad \quad \quad \quad \,\,
H^{2}&=&\frac{\kappa^{2}_{4}\rho}{3} \Big{(} 1-
\frac{6}{\kappa^{2}_{4} r^{2}_{c}\lambda}\Big{)} .
\label{lowapprox} \eea

It was shown, that the gravitational potential in this case,
exhibits a four-dimensional behaviour \cite{ktt} and therefore,
equations (\ref{approxm1})-(\ref{casee-})  describe high energy
corrections to the standard Friedmann equation due to induced
gravity. However, only equations (\ref{approxm1}) and
(\ref{approxm2}) are giving $\rho$ dependent corrections to the
standard Friedmann equation while equations (\ref{casee+}) and
(\ref{casee-}) are giving the standard Friedmann equation with a
redefined energy density or a cosmological constant. Note also
that the last term in (\ref{approxm2}) is giving a small
correction to the second term and therefore cannot be
differentiated from equation (\ref{approxm1}). In the low energy
limit, we obtain again the
standard Friedmann equation with a redefined energy density or with a cosmological constant.\\
\\
2) \,\, \underline{$\mu\ll 1$}.\, In the high energy limit,
(\ref{approxfried}) can be approximated by \bea
\kappa^{2}_{4}r^{2}_{c}\rho \gg1 ,\,\quad \quad \quad \quad
H^{2}&=&\frac{\kappa^{2}_{4}\rho}{3} \Big{(}
1+\epsilon \frac{2\sqrt{3}}{\kappa_{4} r_{c}\rho^{1/2}}\Big{)},\label{approxm12} \\
\kappa^{2}_{4}r^{2}_{c}\rho \ll1 ,\, \epsilon=+1,\,\, H^{2}&=& 2
\frac{\kappa^{2}_{4}\rho}{3}\Big{(} 1+
\frac{6}{\kappa^{2}_{4} r^{2}_{c}\rho}\Big{)},\\
 \epsilon=-1,\,\,
H^{2}&=&\kappa^{4}_{5}\frac{\rho^{2}}{36}\Big{(}1+2\frac{\lambda}{\rho}\Big{)}=
\mu\frac{\kappa^{2}_{4}\rho}{3}\Big{(}1+\frac{\rho}{2\lambda}\Big{)}.
\label{rudsud} \eea Equation (\ref{approxm12}) represents
four-dimensional corrections to the Friedmann equation, while
equation (\ref{rudsud}) represents five-dimensional corrections.
  In the low energy limit, for $\epsilon=-1$ there is a transition \cite{deffayet,ktt} from
five dimensions to four dimensions, while for $\epsilon=+1$ the
Friedmann equation is a constant.

We turn now to a Minkowski bulk where $\Lambda_{5}=0$. In this
case equation (\ref{igr}) becomes
 \be
H^{2}=\frac{\kappa^{2}_{4}\rho}{3} \Big{\{}
1+\frac{6}{\kappa^{2}_{4}
r^{2}_{c}\rho}+\frac{\lambda}{\rho}+\epsilon\,
\frac{2\sqrt{3}}{\kappa_{4} r_{c}\rho^{1/2}}\Big{(}
1+\frac{3}{\kappa^{2}_{4} r^{2}_{c}\rho}+\frac{\lambda}{\rho}
\Big{)}^{1/2}\Big{\}}. \label{approxfriedl5} \ee We can
distinguish the following cases:\\
\\
a)\,\, \underline{$\mu \gg 1$}.\, In the high energy limit
$\rho\gg\lambda$, the Friedmann equation (\ref{approxfriedl5}) is
approximated by  \bea \kappa_{4}r_{c}\rho^{1/2}
\Big{(}\frac{\lambda}{\rho} \Big{)}\ll1 ,\,
&H^{2}&=\frac{\kappa^{2}_{4}\rho}{3} \Big{(}
1+\epsilon\frac{2\sqrt{3}}{\kappa_{4} r_{c}\rho^{1/2}}\Big{)},\label{approxml51} \\
\kappa_{4}r_{c}\rho^{1/2} \Big{(}\frac{\lambda}{\rho} \Big{)}\sim
1 , \, &H^{2}&=\frac{\kappa^{2}_{4}\rho}{3} \Big{(}
1+\frac{\lambda}{\rho}+\epsilon \frac{2\sqrt{3}}{\kappa_{4}
r_{c}\rho^{1/2}}\Big{)}, \label{approxml52}\\
\kappa_{4}r_{c}\rho^{1/2} \Big{(}\frac{\lambda}{\rho} \Big{)}\gg1
,\, &H^{2}&=\frac{\kappa^{2}_{4}\rho}{3} \Big{(}
1+\frac{\lambda}{\rho}\Big{)}. \label{approxml53} \eea Equations
(\ref{approxml51})-(\ref{approxml53}) represent pure
four-dimensional corrections to the standard Friedmann equation.
In the low energy limit we get the usual linearized Friedmann
equation of induced gravity with an effective cosmological
constant \be H^{2}=\frac{\kappa^{2}_{4}\lambda}{3} \Big{(}
1+\frac{\rho}{\lambda}+\epsilon \frac{2\sqrt{3}}{\kappa_{4}
r_{c}\lambda^{1/2}}+\frac{6}{\kappa^{2}_{4}r^{2}_{c}\lambda}\Big{)}. \ee\\
\\
b)\,\,\underline{$\mu \ll 1$}.\, In the high energy limit,
(\ref{approxfriedl5}) can be approximated by \bea
\kappa^{2}_{4}r^{2}_{c}\rho\gg1 ,\,
&H^{2}&=\frac{\kappa^{2}_{4}\rho}{3} \Big{(}
1+\epsilon\frac{2\sqrt{3}}{\kappa_{4} r_{c}\rho^{1/2}}\Big{)},\label{approxml5s1} \\
\kappa^{2}_{4}r^{2}_{c}\rho\ll1,\, \,\epsilon=+1,\,\
&H^{2}&=\frac{4}{r_{c}^{2}} \Big{(} 1+\frac{\kappa^{2}_{4}
r^{2}_{c}\rho}{6}\Big{)}, \label{approxml5s2}\\
\epsilon=-1,\,\
&H^{2}&=\frac{\kappa^{4}_{5}(\rho+\lambda)^{2}}{36}\approx
\mu\frac{\kappa^{2}_{4}\rho}{3}\Big{(}1+\frac{\rho}{2\lambda}\Big{)}.
\label{approxml5s3} \eea Equation (\ref{approxml5s1}) represents
four-dimensional correction to the standard Friedmann equation,
while equation (\ref{approxml5s3}) is a pure five-dimensional
correction. Note that if we ignore the $\lambda^{2}$ term which is
small in the high energy limit, equation (\ref{approxml5s3}) is
identical to equation (\ref{rudsud}). In the low energy limit,
equations (\ref{approxml5s2}) and (\ref{approxml5s3}) go to their
constant limits.

From the previous analysis we can distinguish two different
cosmological evolutions of a brane-universe with induced gravity.
The first is a pure four-dimensional evolution at all energies
(distances) if $\lambda\gg 6M^{6}_{5}/M^{2}_{4}$ ($\mu \gg 1$).
The second is an interesting evolution if $\lambda\ll
6M^{6}_{5}/M^{2}_{4}$ ($\mu \ll 1$). At small energies $E\ll
M^{3}_{5}/M^{2}_{Pl}$ or large energies $E\gg M^{3}_{5}/M^{2}_{4}$
($r\gg 1/ \kappa^{2}_{4}\lambda $ and $r\ll r_{c}$ respectively)
the evolution is four-dimensional, while at an intermediate energy
scale $M^{3}_{5}/M^{2}_{Pl}\ll E \ll M^{3}_{5}/M^{2}_{4}$ ($
6/\kappa^{2}_{5}\lambda \gg r \gg r_{c}$) is five-dimensional. The
same behaviour was found in \cite{ktt} where a different fine
tuning was used, while in \cite{deffayet} the five-dimensional
behaviour was recovered with no brane tension. Both cosmological
evolutions are shown in Fig. 1.
%%%%%%%%%%%%%%%%%%%%%%%%%%%%%%%%%%%%%%%%%%%%%%%%%%%%%%%%%%%%%
\begin{figure}[t]
\centering
\hspace{0.1cm}%
\includegraphics[width=.85\textwidth]{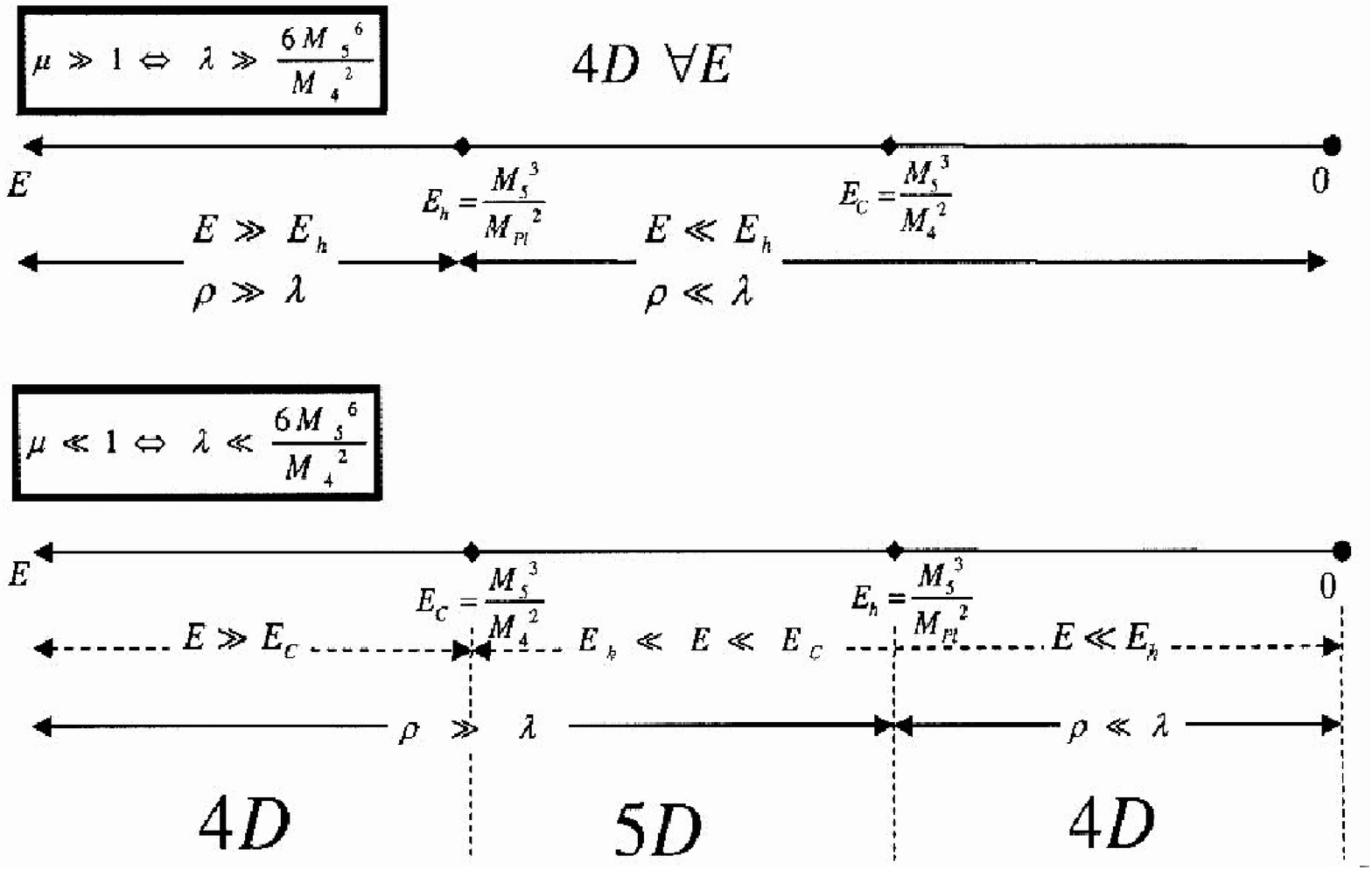}
 \caption{The cosmological evolution of a brane-universe with
 induced gravity.}
\end{figure}
%%%%%%%%%%%%%%%%%%%%%%%%%%%%%%%%%%%%%%%%%%%%%%%%%%%%%%%%%%%%%%%%%%%

In the high energy limit therefore, if $E\gg M^{3}_{5}/M^{2}_{4}$
we have corrections to the four-dimensional Friedmann equation,
due to induced gravity, given by \be
H^{2}=\frac{\kappa^{2}_{4}\rho}{3} \Big{(}
1+\epsilon\frac{2\sqrt{3}}{\kappa_{4} r_{c}\rho^{1/2}}\Big{)},
\label{approxfriedoup} \ee and if $M^{3}_{5}/M^{2}_{Pl}\ll E \ll
M^{3}_{5}/M^{2}_{4}$ we have corrections to the five-dimensional
Friedmann equation of the Randall-Sundrum type \be
H^{2}=\kappa^{4}_{5}\frac{\rho^{2}}{36}\Big{(}1+2\frac{\lambda}{\rho}\Big{)}.
\label{secondcorr} \ee
 In the next section, we will apply the slow-roll
approximation of the inflationary scenario, to the two high energy
corrections of the Friedmann equation given by
(\ref{approxfriedoup}) and (\ref{secondcorr}).

\section{Inflationary dynamics on the brane}

We will assume that the inflationary dynamics is parametrized by a
scalar field $\phi$ with a self-interaction potential $V(\phi)$,
which is confined on the brane and dominates the energy momentum
tensor $T_{\mu \nu } $ on the brane. Then it satisfies the
Klein-Gordon equation \be
\ddot{\phi}+3H\dot{\phi}+V^{\prime}(\phi)=0, \ee since
$\nabla^{\mu}T_{\mu\nu}=0$ on the brane. This is equivalent to the
continuity equation $ \dot{\rho}+3H\rho(1+w)=0$, where we have
assume an equation of state $p=w\rho$, with $\rho$ and p given by
\bea
\rho&=&\frac{1}{2} \dot{\phi}^{2}+V(\phi),\\
p&=&\frac{1}{2} \dot{\phi}^{2}-V(\phi). \eea We are interested on
the range of parameter space where inflation potential dominates
the brane tension and assume the slow-roll approximation, $
\dot{\phi}^{2}\ll V$ and $|\ddot{\phi}|\ll H |\phi|$, and
therefore $\rho\approx V$. These assumptions can be justified on
the grounds that as we showed in the previous section, induced
gravity introduces corrections to the Friedmann equation.

The inflationary parameters $\varepsilon=-\dot{H}/H^{2}$ and
$\eta=V''/3H^{2}$ with the use of (\ref{igr}), (\ref{secondfried})
and (\ref{finetune}) become \bea \varepsilon &=&
\mu\Big{(}\frac{1}{2\kappa^{2}_{Pl}} \frac{V^{\prime
2}}{V^{2}}\Big{)} \Big{[} \frac{
1+\epsilon\,\,2\sqrt{3}\Big{(}A^{2}+4\kappa_{4}^{2}
r_{c}^{2}V\Big{)}^{-1/2}} {\Big{(} 1+\frac{6}{\kappa^{2}_{4}
r^{2}_{c}V}+\frac{\lambda}{V}+\epsilon\,
\frac{\sqrt{3}}{\kappa_{4}^{2} r_{c}^{2}V}
\Big{(}A^{2}+4\kappa_{4}^{2}
r_{c}^{2}V\Big{)}^{1/2}\Big{)}^{2}}\Big{]}, \label{epsilon}\\
\eta&=&\mu\Big{(}\frac{1}{\kappa^{2}_{Pl}}\frac{V''}{V} \Big{)}
\Big{[}\frac {1}
 { 1+\frac{6}{\kappa^{2}_{4}
r^{2}_{c}V}+\frac{\lambda}{V}+\epsilon\,
\frac{\sqrt{3}}{\kappa_{4}^{2}
r_{c}^{2}V}\Big{(}A^{2}+4\kappa_{4}^{2} r_{c}^{2}V\Big{)}^{1/2}
}\Big{]},\label{hta} \eea where $A^{2}=(4\kappa^{2}_{4}
r^{2}_{c}\lambda+12 -2\Lambda_{5}r^{2}_{c})$ and
$\mu=\kappa^{2}_{Pl}/\kappa^{2}_{4}$ as it will be discussed
later. The term in the curved brackets corresponds to the GR
inflationary parameter $\varepsilon$. Hence we can see that for
the cosmological evolution with $\mu \ll 1$ induced gravity
effects ease for a given potential the condition for slow-roll
inflation which is $max\{\varepsilon,|\eta|\} \ll 1$. The number
of e-folds during inflation is given by
$N=\int^{t_{f}}_{t_{i}}Hdt$ which in the slow-roll approximation
using (\ref{igr}) becomes \bea
N&=&-\kappa^{2}_{4}\int^{\phi_{f}}_{\phi_{i}}\frac{V}{V^{\prime}}d\phi
-\kappa^{2}_{4}\int^{\phi_{f}}_{\phi_{i}}\frac{1}{V^{\prime}}
\Big{(}\lambda +\frac{6}{\kappa_{4}^{2} r_{c}^{2}}
\Big{)}d\phi\nonumber \\
&-&\epsilon\frac{\sqrt{3}}{r^{2}_{c}}\int^{\phi_{f}}_{\phi_{i}}\frac{1}{V^{\prime}}
\Big{(}4\kappa_{4}^{2} r_{c}^{2}V+A^{2}   \Big{)}^{1/2} d\phi,
\label{efolgrgen}\eea
 where $\phi_{i}$ and $\phi_{f}$ are the values of the
inflaton field at the beginning and at the end of inflation. The
scalar perturbations in the slow-roll approximation are given by
\be A^{2}_{s}=
\frac{1}{\mu^{3}}\Big{(}\frac{\kappa^{6}_{Pl}}{75\pi^{2}}
\frac{V^{3}}{V^{\prime 2}}\Big{)} \Big{[}
1+\frac{6}{\kappa^{2}_{4} r^{2}_{c}V}+\frac{\lambda}{V}+\epsilon\,
\frac{\sqrt{3}}{\kappa_{4}^{2} r_{c}^{2}V}
\Big{(}A^{2}+4\kappa_{4}^{2} r_{c}^{2}V\Big{)}^{1/2}\Big{]}^{3}
.\label{spgrgen} \ee We cannot treat the inflationary parameters $
\epsilon$ and $\eta$ of (\ref{epsilon}) and (\ref{hta}) and also
equations (\ref{efolgrgen}) and (\ref {spgrgen}) analytically due
to their complexity. We will analyse instead the inflationary
dynamics which arises from the Friedmann equations
(\ref{approxfriedoup}) and (\ref{secondcorr}).

We start first with the Friedmann equation (\ref{approxfriedoup}).
The condition for inflation is \be
\frac{\ddot{a}}{a}=H^{2}+\dot{H}=\frac{\kappa^{2}_{4}}{3}\Big{(}
V\Big{(}1+\epsilon\frac{2\sqrt{3}}{\kappa_{4}
r_{c}V^{1/2}}\Big{)}-\dot{\phi}^{2}\Big{(}1+\epsilon\frac{\sqrt{3}/2}{\kappa_{4}
r_{c}V^{1/2}}\Big{)}\Big{)}>0, \ee which reduces to \be
\dot{\phi}^{2}\Big{(}1+\epsilon\frac{\sqrt{3}/2}{\kappa_{4}
r_{c}V^{1/2}}
\Big{)}<V\Big{(}1+\epsilon\frac{2\sqrt{3}}{\kappa_{4}
r_{c}V^{1/2}} \Big{)}. \label{condition} \ee The above condition
(\ref{condition}) coincides with the GR condition for inflation
when $\kappa_{4} r_{c}V^{1/2}\gg 1$.

 Using
(\ref{approxfriedoup}) the slow-roll parameters become \bea
\varepsilon &=& \mu\Big{(}\frac{1}{2\kappa^{2}_{Pl}}
\frac{V^{\prime 2}}{V^{2}}\Big{)}
\Big{(}1-\epsilon\frac{3\sqrt{3}} {\kappa_{4}
r_{c}V^{1/2}}\Big{)},\label{eparam} \\
\eta&=&\mu\Big{(}\frac{1}{\kappa^{2}_{Pl}}\frac{V''}{V}
\Big{)}\Big{(} 1-\epsilon\frac{2\sqrt{3}}{\kappa_{4}
r_{c}V^{1/2}}\Big{)}. \label{eta1} \eea Similarly, the number of
e-folds during inflation and the scalar perturbations become \bea
N&=&-\frac{\kappa^{2}_{Pl}}{\mu}\int^{\phi_{f}}_{\phi_{i}}\frac{V}{V^{\prime}}
\Big{(}1+\epsilon\frac{2\sqrt{3}}{\kappa_{4} r_{c}V^{1/2}}\Big{)}
d\phi,\label{efolgr}\\
A^{2}_{s}&=&
\frac{1}{\mu^{3}}\Big{(}\frac{\kappa^{6}_{Pl}}{75\pi^{2}}
\frac{V^{3}}{V^{\prime 2}}\Big{)} \Big{(}
1+\epsilon\frac{6\sqrt{3}}{\kappa_{4}
r_{c}V^{1/2}}\Big{)}.\label{spgr}  \eea
 Equations (\ref{eparam})-(\ref{spgr}) are representing high
 energy corrections to the general relativity (GR)
 corresponding quantities in the slow-roll approximation, and coincide
 with them in the limit $r_{c}\rightarrow\infty$.

Considering a quadratic potential $V=V_{0}\phi^{2}$, in the GR
limit $r_{c}\rightarrow\infty$, the spectral index of the scalar
perturbation spectrum is independent of $V_{0}$ and it is given by
$n_{s}=1+2\eta-6\varepsilon=1-8/\kappa^{2}_{4}\phi^{2}$. Assuming
$\phi_{f}\sim 10^{-1}\phi_{i}$, from equations (\ref{efolgr}) and
(\ref{spgr}) the values of $V_{0}$ and $\phi_{i}$ can be
calculated. Using the COBE normalization for the scalar
perturbations $A^{2}_{s}=4\times10^{-10}$, and for the number of
e-folds $N=55$, we find $V_{0}=3.46\times10^{27}\,(GeV)^{2}$ and
$\phi_{i}=1.79\times10^{20}\,GeV$, while for $N=70$ the numbers
are $V_{0}=2.15\times10^{27}\,(GeV)^{2}$ and
$\phi_{i}=2.02\times10^{20}\,GeV $ respectively. These values of
$V_{0}$ correspond to an inflaton mass of $\sim 10^{13}\,GeV$, and
the values of the inflaton field are larger than $3M_{4}$. The
spectral index of the scalar perturbation spectrum for $N=55$ is
$n_{s}=0.9640$, while for $N=70$ we get a larger value of
$n_{s}=0.9716$. This is known as the chaotic inflationary scenario
\cite{linde}, and it was criticised for its super-Planckian field
values. The problem with the super-Planckian field values is that
one expects non-renormalizable quantum corrections to completely
dominate the potential, destroying the flatness of the potential
required for inflation. This is known as the $\eta$-problem
\cite{lyth}.

Coming back to the induced gravity corrections, equation
(\ref{efolgr}) for a quadratic potential becomes \be
N=\frac{\kappa^{2}_{4}\phi^{2}}{4}(1-c^{2})+\epsilon
\sqrt{3}\frac{\kappa_{4}\phi}{r_{c}V^{1/2}_{0}}(1-c), \ee where
$\phi=\phi_{i}$ and $c=\phi_{f}/\phi_{i}$. From this equation, the
crossover scale can be calculated, \be r_{c}=\epsilon
\frac{\sqrt{3}\kappa_{4}\phi(1-c)}{V_{0}^{1/2}}\Big{(}\frac{4}{4N-\kappa^{2}_{4}
\phi^{2}(1-c^{2})}\Big{)}. \label{scale} \ee To obtain a positive
five-dimensional mass $M_{5}$, considering also the GR value of
$c\sim 10^{-1}$, equation (\ref{scale}) gives the following
conditions for the square of the inflaton field \bea
\epsilon&=&+1,\,\,\ \phi^{2}<\frac{4N}{\kappa^{2}_{4}(1-c^{2})},\label{phipos}\\
\epsilon&=&-1,\,\,\
\phi^{2}>\frac{4N}{\kappa^{2}_{4}(1-c^{2})}\label{phineg}. \eea
The spectral index $n_{s}=1+2\eta-6\varepsilon$ using
(\ref{eparam}) and (\ref{eta1}) becomes \be
n_{s}=1-\frac{4}{\kappa^{2}_{4}\phi^{2}}\Big{(}2-\epsilon\frac{7\sqrt{3}}{\kappa_{4}
r_{c}\sqrt{V_{0}}\phi}\Big{)}.\label{spectindex} \ee Substituting
(\ref{scale}) into (\ref{spectindex}) we can express the inflaton
field as a function of the spectral index \be
\phi_{\pm}^{2}=\frac{1}{2\kappa^{2}_{4}(1-n_{s})}\Big{(}(15+7c)\pm\Big{(}
\frac{(15+7c)^{2}-112N(1-n_{s})}{(1-c)} \Big{)}^{1/2}\Big{)}.
\label{phi1} \ee The inflaton field is well defined if \be
n_{s}>1- \frac{(15+7c)^{2}(1-c)}{112N}. \label{nscon} \ee From
this equation, using also (\ref{phipos}) and (\ref{phineg}) which
are the positivity conditions of the crossover scale, we find that
the crossover scale is always negative for the solution
($\epsilon=-1,\,\ \phi_{-}$), for the solution ($\epsilon=+1,\,\
\phi_{-}$) is always positive and there is a lower bound for the
spectral index given by (\ref{nscon}), while for the other
solutions we find the following lower or upper bounds for the
spectral index $n_{S}$ \bea \epsilon=+1, \quad \phi_{+}, \quad
              &n_{s}&<1-\frac{(15+7c)^{2}(1-c)}{112N},
              \\
              \epsilon=-1,  \quad  \phi_{+}, \quad
              &n_{s}&>1-\frac{2(1-c^{2})}{N}\label{posconstr}.
\eea For $c=10^{-1}$ and for the number of e-folds $N=55$ and
$N=70$, the critical value of the spectral index is $n_{s}=0.9640$
and $n_{s}=0.9717$ respectively. Therefore, for the  solution
having ($\epsilon=+1,\,\ \phi_{+}$) we do not expect any
improvement of the value of the spectral index from its GR limit.

 Using the COBE normalization for the
scalar perturbations $A^{2}_{s}=4\times10^{-10}$, equation
(\ref{spgr}), with the use of (\ref{scale}), can give $V_{0}$ as a
function of $\phi$ \be V_{0}=\frac{3\pi^{2}(1-c)}
{125\times10^{5}\kappa^{4}_{4}\phi^{2}\Big{(}12N-\kappa^{2}_{4}
\phi^{2}(1-c)(1+3c)\Big{)}}. \label{vo}\ee Demanding a positive
$V_{0}$ we get further constraints for the spectral index for the
$\phi_{+}$ solution \be \epsilon=\pm 1,\,\,\ \phi_{+},\,\,\,
n_{s}<1-\frac{19(1-c)(1+3c)}{18N},\label{limit+}
 \ee while for the $\phi_{-}$ it is always positive. For
$c=10^{-1}$ and for the number of e-folds $N=55$ and $N=70$, the
critical value of the spectral index for these constraints is
$n_{s}=0.9775$ and $n_{s}=0.9824$ respectively. Therefore, for the
allowed solutions from the positivity constraints, solution
($\epsilon=+1,\,\ \phi_{-}$) has a lower bound of the spectral
index but no upper one, while for the solution ($\epsilon=-1,\,\
\phi_{+}$) from (\ref{posconstr}) and (\ref{limit+}) the spectral
index is limited to $0.9640<n_{s}<0.9775$ for the number of
e-folds $N=55$, and to $0.9717<n_{s}<0.9824$ for the number of
e-folds $N=70$.

We can verify these results if we substitute the inflaton field
(\ref{phi1}) into (\ref{vo}). Then, Fig. 2 shows the variation of
the spectral index for various values of $V_{0}$, after having
substituting the $\phi_{+}$ solution into (\ref{vo}). There is an
upper bound of $n_{s}$ in agreement with (\ref{limit+}). In Fig. 3
  we plot the variation of the spectral index for the $\phi_{-}$ solution.
According to positivity constraints there is no upper bound for
the spectral index. Finally, for various values of $V_{0}$
 the inflaton field can be calculated and
   from the crossover scale (\ref{scale}), the five-dimensional mass $M_{5}$ can be
specified.

Coming now to the five-dimensional corrections, the Friedmann
equation (\ref{secondcorr}) can be written as \be H^{2}=\mu
\frac{\kappa^{2}_{4}\rho}{3}\Big{(}1+\frac{\rho}{2\lambda}\Big{)}.\label{indrudsud}
\ee The $\mu$ parameter which appears in front of $\kappa^{2}_{4}$
also appears in the low energy limit of the Friedmann equation. In
order to recover the four dimensional Newton's constant of early
cosmology we have to define \be \mu\kappa^{2}_{4}=\kappa^{2}_{Pl}.
\label{newtconst} \ee Performing the same analysis of the
inflationary dynamics as before, with the Newton's constant
appearing in (\ref{indrudsud}) identified with the standard
four-dimensional Newton's constant using (\ref{newtconst}), we
find for $N=55$ the spectral index $n_{s}=0.9548$, while for
$N=70$ the spectral index becomes $n_{s}=0.9644$ which are the
Randall-Sundrum values of the spectral index \cite{liddle}. The
redefinition (\ref{newtconst}) has important consequences in the
early universe. Because $\mu$ is a small number, it gives
$G_{4}>G_{Pl}$ which makes gravity stronger \cite{ktt} in the high
energy limit \cite{wands}.

%%%%%%%%%%%%%%%%%%%%%%%%%%%%%%%%%%%%%%%%%%%%%%%%%%%%%%%%%%%%%%%%%%%%
\begin{figure}[h]
\centering \hspace{0.1cm}
\begin{tabular}{cc}
\includegraphics[scale=.6]{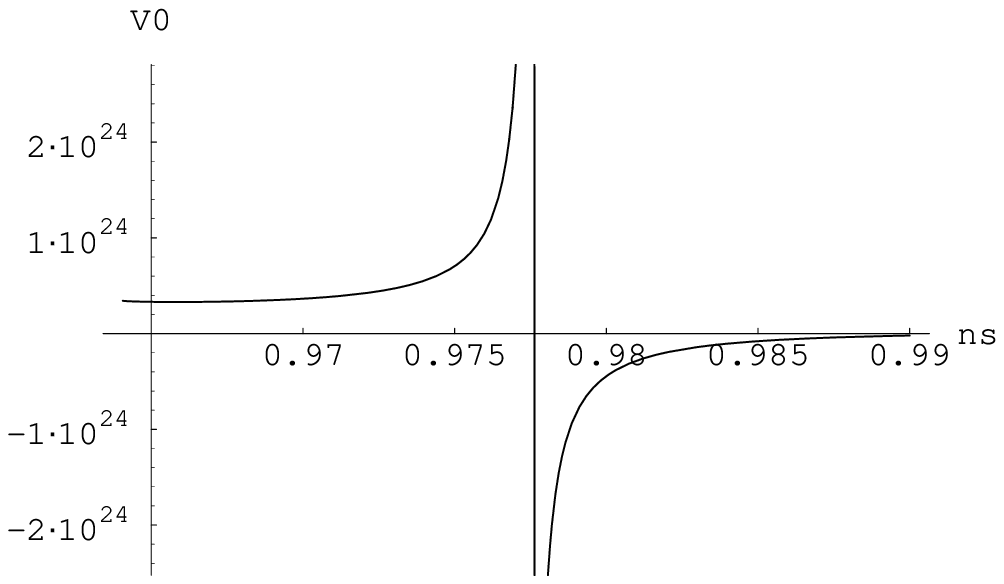}&
\hspace{0.1cm}
\includegraphics[scale=.6]{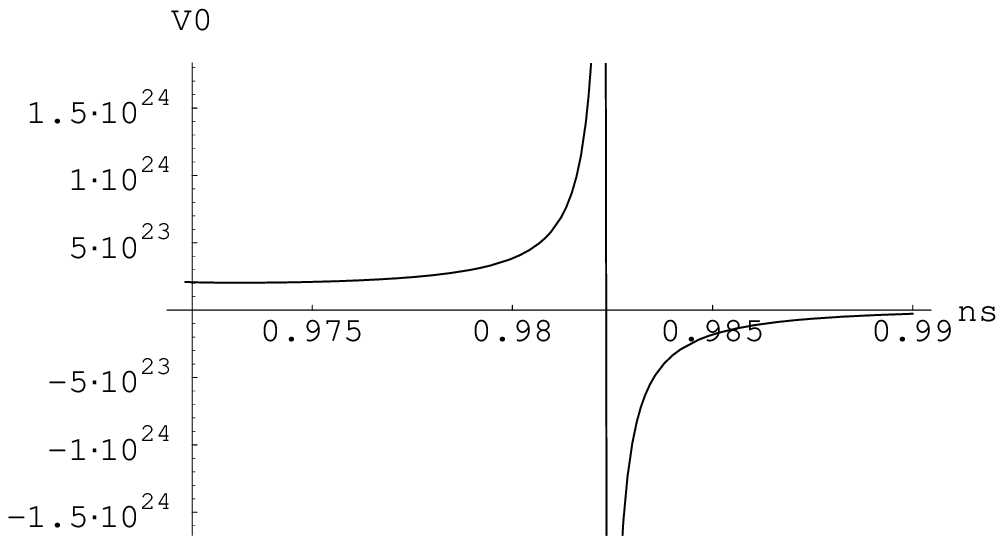}\\
%(a): $\zeta >0$, $\beta=0$&(b):  $\zeta <0$, $\beta=0$
\end{tabular}
 \caption{
 $V_{0}$ versus $n_{s}$ for the solution $\phi_{+}$ and for the
number of e-folds $N=55$ and $N=70$ respectively. }
\end{figure}
%%%%%%%%%%%%%%%%%%%%%%%%%%%%%%%%%%%%%%%%%%%%%%%%%%%%%%%%%%%%%%%%%%%%%

%%%%%%%%%%%%%%%%%%%%%%%%%%%%%%%%%%%%%%%%%%%%%%%%%%%%%%%%%%%%%%%%%%
\begin{figure}[h]
\centering
\hspace{0.1cm}%
\begin{tabular}{cc}
\includegraphics[scale=.6]{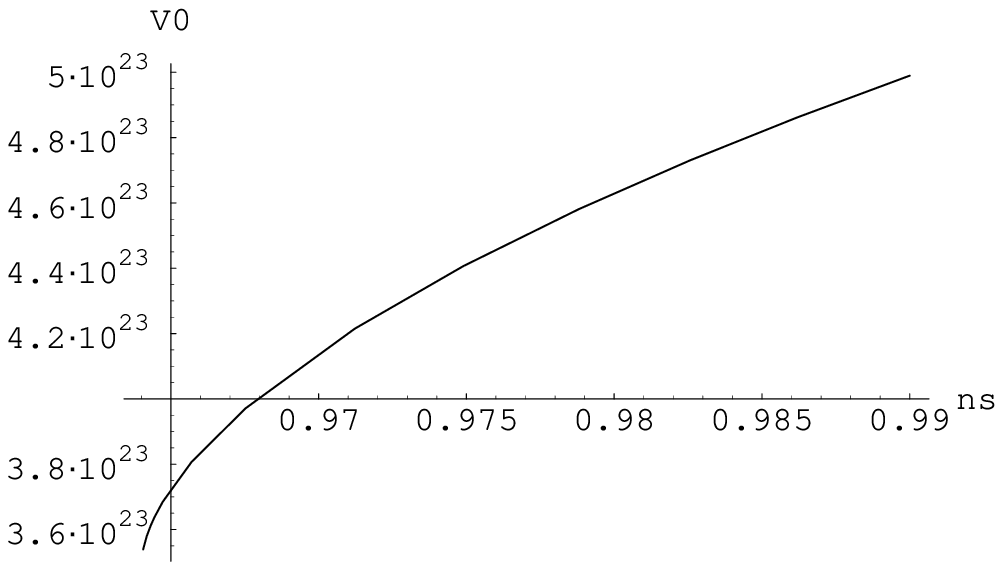}&
\hspace{0.1cm}
\includegraphics[scale=.6]{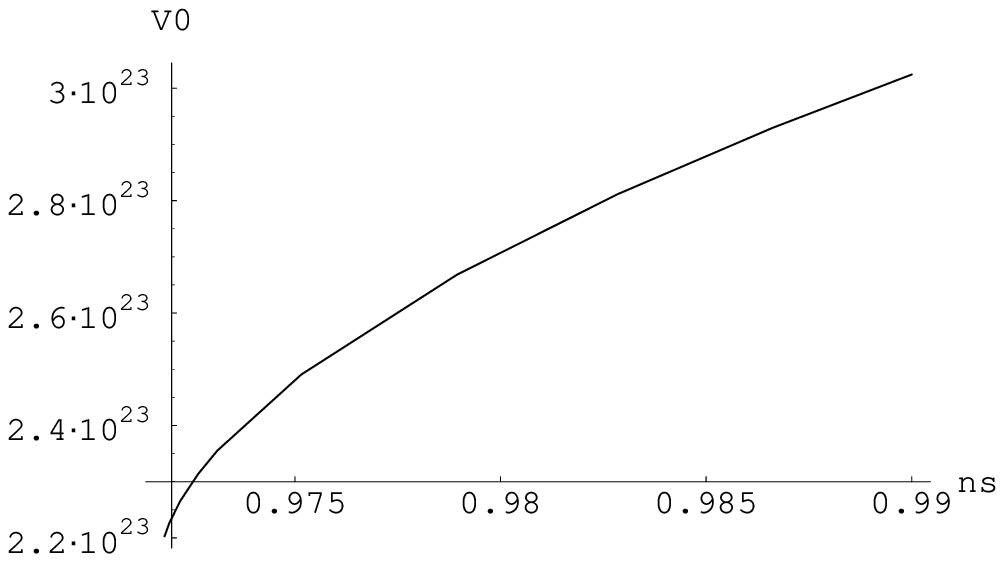}\\
%(a): $\zeta >0$, $\beta=0$&(b):  $\zeta <0$, $\beta=0$
\end{tabular}
 \caption{
$V_{0}$ versus $n_{s}$ for the solution $\phi_{-}$ and for the
number of e-folds $N=55$ and $N=70$ respectively.}
\end{figure}
%%%%%%%%%%%%%%%%%%%%%%%%%%%%%%%%%%%%%%%%%%%%%%%%%%%%%%%%%%%%%%%%%%%%%

\section{Chaotic braneworld models with induced gravity}

In this section we will search for realistic chaotic inflationary
models of induced gravity describing the early evolution. As
showed in Fig. 1, for $\lambda\gg 6M_{5}^{6}/M^{2}_{4}$ we have a
four-dimensional cosmological evolution for all energies, while if
$\lambda\ll 6M_{5}^{6}/M^{2}_{4}$ a four-dimensional evolution is
followed by a five-dimensional at high energies. The
five-dimensional mass $M_{5}$ in all cases considered in Sec. III,
is of the order $\sim 10^{17}\,\, GeV$. Hence, the first case of
$\lambda \gg 6M_{5}^{6}/M^{2}_{4}$ is excluded, because we probe a
high energy regime where quantum gravity corrections become
important.

We studied in Sec. II the cosmological evolution of models with
AdS or Minkowski bulk by setting the effective cosmological
constant or the five-dimensional cosmological constant equal to
zero respectively. In both cases for $\lambda \ll
6M_{5}^{6}/M^{2}_{4}$, and at energies $E\gg M_{5}^{3}/M^{2}_{4}$
the cosmological evolution is described by the Friedmann equation
(\ref{approxfriedoup}). For $\rho \gg \lambda$ and in the energy
regime $M^{3}_{5}/M^{2}_{Pl}\ll E \ll M^{3}_{5}/M^{2}_{4}$ the
Friedmann equation is given by (\ref{secondcorr}). In the case of
an AdS bulk and for $\epsilon=+1$ the brane tension is negative.
 In general, we expect a negative brane
tension to give at late times a negative Newton's constant.

This is the case in the simplest one-brane Randall-Sundrum model.
In this model, the effective gravitational constant in the late
time cosmology is identified with the Newton's constant and a
simple relation between the brane tension $\lambda$ the
five-dimensional mass $M_{5}$ and the $G_{Pl}$ results. The
situation in induced gravity is more subtle. In \cite{mmt} it was
shown that in late time cosmological evolution and without any
fine-tuning, the effective gravitational constant is a complicated
function of the parameters, and it is changing in time. In our
case for the fine-tuning (\ref{finetune}) and the approximations
we used in Sec. II, in the late time cosmology the Newton's
constant is fixed by relation (\ref{newtconst}).

Negative tension branes appear quite naturally as endpoints of
spacetime in higher than four dimensions. For example they may
appear in non-oriented string theories as orientifold planes
\cite{polsinski}. In Randall-Sundrum model their presence provides
a solution to the hierarchy problem \cite{randall} and it appears
that they are needed to any realization of geometrical hierarchy
in string theories \cite{polsisk}. However, the presence of
negative tension branes signals an instability in the bulk-brane
system \cite{charmousis1}. What really happens is that tachyonic
modes are generated via tensor perturbations of the spacetime and
bound to the negative tension brane much like the graviton
zero-mode to the positive tension brane.

There are various ways to stabilize an unstable bulk-brane system.
One way is to introduce a scalar field in the bulk which has
self-interaction potentials at the branes. This is the way one can
stabilise the two-brane Randall-Sundrum model \cite{goldberger}.
Another way is to add an effective cosmological constant. This
corresponds to a system rolling down a potential hill towards a
different vacuum of the theory. Actually the reason why we find a
negative brane tension is that we have put the effective
cosmological constant equal to zero in the first place. Finally,
one can introduce additional terms in the action in such a way as
to drive the system to a stable vacuum.

In the last section we showed that the two solutions
($\epsilon=+1,~\phi_{-}$) and ($\epsilon=-1,~\phi_{+}$) improve
the GR value of the spectral index. However, the solution
($\epsilon=+1,~\phi_{-}$) gives the unexpected result of an
unbounded from the above spectral index as it is shown in Fig. 3.
This is happening because for this solution the inflation never
ends. Inflation ends when the slow-roll parameter $\varepsilon=1$.
Substituting a quadratic potential $V=V_{0}\phi^{2}$ into
(\ref{eparam}) we get \be
\varepsilon=\frac{2}{\kappa^{2}_{4}\phi^{2}_{f}}\Big{(}1-\epsilon\frac{3\sqrt{3}}
{\kappa_{4}r_{c}\sqrt{V_{0}}\phi_{f}}\Big{)}. \label{eps-}\ee For
the solution ($\epsilon=+1,~\phi_{-}$) using the GR value
$\phi_{f}=10^{-1}\phi_{i}$ we find that the condition at the end
of the inflation $\varepsilon=1$ is not satisfied. This is
happening because of the negative sign of the second term in
(\ref{eps-}).

For $E \gg M_{5}^{3}/M^{2}_{4}$ we are left with the solution
($\epsilon=-1, \phi_{+}$). This solution can have positive brane
tension for both AdS and Minkowski bulk, well defined energy
scales and it gives a spectral index closer to the observations.
If we take the GR value of $V_{0}=3.46\times 10^{27}\, (GeV)^{2}$
for the number $N=55$ of e-folds, for $\mu = 0.01$ the value of
the inflaton field is $2.85\times10^{18}\,GeV$. The
five-dimensional mass is $M_{5}=1.24\times10^{16}\,GeV$ and it is
less than the Planck mass. Therefore we see, that here we do not
have the standard problem of the chaotic inflation. The value of
the inflaton field is less than the Planck scale, larger however
than the five-dimensional mass, and hence we do not expect to have
uncontrollable  quantum gravity corrections. The reason is that
gravity becomes stronger in early cosmological evolution bringing
down the value of the inflaton field, as can be seen in
(\ref{newtconst}). The spectral index is $n_{s}=0.9775$ better
than the GR value of $n_{s}=0.9640$. If we take $N=70$ for the
number of e-folds than the spectral index becomes $n_{s}=0.9824$.
In this chaotic inflationary model of the induced gravity the
initial value of the potential is not fixed as in the GR limit.
The maximum value it can take is
$V_{0}^{1/2}=m_{\phi}=10^{-5}M_{P}$ which for
$M_{P}=1.2\times10^{19}\,GeV$ is $V_{0}=1.44\times 10^{28}\,
(GeV)^{2}$. Then, with this value of $V_{0}$ the inflaton field
and the five-dimensional mass change slightly, so does the
spectral index which has already reached its maximum allowed
value.

For an energy regime where $M^{3}_{5}/M^{2}_{Pl}\ll E \ll
M^{3}_{5}/M^{2}_{4}$ the cosmological evolution is governed by
(\ref{indrudsud}). In general this Friedmann equation is giving a
different cosmological evolution than the Randall-Sundrum model.
However, in the high energy limit $\rho\gg \lambda$, considering
the induced gravity as correction to the Randall-Sundrum model, we
fix the Newton's constant from (\ref{newtconst}) and then we
recover the known Randall-Sundrum model results. It is remarkable
that the same result applies also to a Minkowski bulk, as this was
noticed in \cite{deffayet} in the case of $\Lambda_{5}=0$ and
$\lambda=0$.

\section{Conclusions}

We studied the early time cosmological evolution of a braneworld
model with induced gravity when the brane tension and the bulk
cosmological constant is included. For $E\gg M_{5}^{3}/M^{2}_{4}$
the evolution is four-dimensional and the Friedmann equation is
the standard Friedmann equation of GR supplemented with correction
terms due to induced gravity. For $M^{3}_{5}/M^{2}_{Pl}\ll E \ll
M^{3}_{5}/M^{2}_{4}$ the evolution is five-dimensional and the
Friedmann equation is of the Randall-Sundrum type. In late times
and at energies $E\ll M_{5}^{3}/M^{2}_{4}$ the evolution is
four-dimensional and the Friedmann equation is the standard
Friedmann equation of GR.

We applied the slow-roll inflationary formalism to this model and
investigated chaotic inflationary scenarios with AdS or Minkowski
bulk. In the induced gravity phase of $E\gg M_{5}^{3}/M^{2}_{4}$,
we found better agreement with the recent observations. The
spectral index of the scalar perturbation spectrum is
$n_{s}=0.9775$ for the number of e-folds $N=55$, while it becomes
$n_{s}=0.9824$, for $N=70$. In the five-dimensional phase we
recover the Rundall-Sundrum model with the known results of the
spectral index of the scalar perturbation spectrum.

An important issue is whether the spectrum of tensor perturbations
is altered in the presence of induced gravity, and also what is
the relation between the scalar and tensor perturbations. The
calculation of tensor perturbations is more involved than the
corresponding calculations in the Rundall-Sundrum model
\cite{mdl}. Up to now it is not known what is the spectrum of
tensor perturbations in braneworls with induced gravity.

This work can be extended to a braneword cosmological model where
both corrections are included in the five-dimensional action: a
four-dimensional scalar curvature from induced gravity on the
brane, and a five-dimensional Gauss-Bonnet curvature term. In
\cite{papa} it was shown, that the combined effect of induced
gravity and the Gauss-Bonnet term, is to smooth out the initial
singularity that is encountered in the individual theories. The
resulted model may improve further the value of the spectral index
of the scalar perturbation spectrum. It may also help to stabilize
the bulk-brane system in the case of a negative tension brane.
\\
\\
{\bf{Note added:}} The paper \cite{zhang} recently appeared which
studies the slow-roll inflationary dynamics in the
five-dimensional RS sector of the induced gravity model.

\section*{Acknowlegements}

We thank K. Dimopoulos, G. Kofinas,  R. Maartens and D. Wands for
valuable discussions and comments. Work supported by the Greek
Education Ministry research program "Hraklitos".

\end{document}